\documentclass[11pt]{article}

\usepackage{amsmath}
\usepackage{amssymb}
\usepackage{graphicx}
\usepackage{cite}
\usepackage{hyperref}

\title{Quantum Bootstrap Approach to a Non-Relativistic Potential for Quarkonium systems}

\author{Jairo Alexis Lopez and Carlos Sandoval \\ Departamento de Física, Universidad Nacional de Colombia, Bogotá, Colombia}

\begin{document}

\maketitle

\begin{abstract}
The quantum bootstrap method is applied to determine the bound-state spectrum of Quarkonium systems using a non-relativistic potential approximation. The method translates the Schrödinger equation into a set of algebraic recursion relations for radial moments $\langle r^m \rangle$, which are constrained by the positive semidefiniteness of their corresponding Hankel matrices. The numerical implementation is first validated by calculating the $1S$ and $1P$ mass centroids for both charmonium ($c\bar{c}$) and bottomonium ($b\bar{b}$) systems, finding deviations of less than 0.5\% from experimental data from the Particle Data Group (PDG). This analysis is then extended to the hypothetical toponium ($t\bar{t}$) system, predicting a $1S$ ground state mass of $M \approx 344.3 \text{ GeV}$. This theoretical mass is in agreement with the energy of the recently observed resonance-like enhancement in the $t\bar{t}$ cross-section by the ATLAS and CMS collaborations. This result provides theoretical support for the interpretation of this experimental phenomenon as the formation of a quasi-bound toponium state and highlights the predictive power of the non-relativistic potential approach for systems of two massive quarks.
\end{abstract}

\newpage
\section{Introduction}
\label{sec:introduction}

The family of mesons known as quarkonium---bound states of a heavy quark and its own antiquark, such as charmonium ($c\bar{c}$) and bottomonium ($b\bar{b}$)---constitutes an exceptional laboratory for probing the theory of the strong interaction, Quantum Chromodynamics (QCD) \cite{GrossWilczek1973, peskin}. The large mass of the constituent quarks ensures that their motion within the bound state is largely non-relativistic, allowing their complex dynamics to be modeled by the non-relativistic Schrödinger equation with an effective potential \cite{peskin}:
\begin{equation}\label{cornell}
V(r)=-\frac{4}{3}\frac{\alpha_s}{r}+\kappa r.
\end{equation}
This approach provides a crucial and tractable bridge between the full relativistic field theory of QCD and the rich, experimentally observed particle spectra.

This potential is a sum of a Coulomb-like term, arising from one-gluon exchange, and a linearly increasing term, reflecting the formation of a confining gluonic flux tube \cite{cornell}. It would account for two of the most fundamental features of the strong force: asymptotic freedom at short distances and color confinement at long distances. While conceptually elegant, accurately solving the Schrödinger equation for this potential has traditionally required approximation techniques or extensive numerical simulations, each with inherent limitations \cite{cornell}.

Traditional methods for solving such quantum systems, like perturbation theory, often fail to provide precise results in the strongly coupled or non-perturbative regimes characteristic of QCD \cite{peskin, cornell}. In recent years, the quantum bootstrap method has emerged as a powerful and innovative non-perturbative framework capable of providing rigorous numerical results for quantum spectra without explicitly solving differential equations\cite{han2020, berenstein2021}. The bootstrap philosophy, which dates back to the S-matrix program of the 1960s, has seen a modern renaissance, fueled by new theoretical insights and computational advances, with success in conformal field theory (CFT) and S-matrix theory \cite{han2020, huang, hu2022}.

The application of this methodology to non-relativistic quantum mechanics is a more recent development\cite{berenstein2021, hu2022,kundu2021, berensteinR}. It relies on fundamental axiomatic principles---symmetry, consistency, and positivity---inherent to any valid quantum theory. These principles are translated into a set of algebraic recursion relations among expectation values of physical observables, combined with powerful positivity constraints on matrices of these moments \cite{hu2022,kundu2021}.This work applies the quantum bootstrap to the non-relativistic quarkonium potential, demonstrating its efficacy in a domain of central importance to particle physics.
The novelty of this approach lies in its non-perturbative robustness and computational efficiency. By leveraging fundamental consistency conditions, the bootstrap method provides a new pathway to determine quarkonium spectra with high precision. The results for the charmonium and bottomonium systems are presented, comparing them against experimental data from the Particle Data Group (PDG)\cite{pdg}.

Furthermore, this analysis is extended to the frontier of the Standard Model: the toponium system. The top quark is unique; its extremely short lifetime ($\tau_t \approx 5 \times 10^{-25}$ s) is an order of magnitude less than the hadronization timescale ($\tau_{\text{had}} \approx 3 \times 10^{-24}$ s), precluding the formation of a stable bound state. However, recent experimental results from the CMS \cite{cms2025} and ATLAS \cite{atlas2025} collaborations at the LHC have provided what could be evidence for a "quasi-bound" toponium state, manifesting as a cross-section enhancement near the $t\bar{t}$ production threshold. The theoretical bootstrap calculation for a hypothetical stable toponium provides a direct prediction for the mass of this observed resonance. 

This paper is organized as follows: Section \ref{sec:formalism} details the quantum bootstrap framework. Section \ref{sec:algorithm} presents the details of the numerical implementation and section \ref{sec:results} presents the results for the proposed potential. Finally, Section \ref{sec:conclusion} offers some conclusions and perspectives for future research.

\section{Formalism and Moment Recursions}
\label{sec:formalism}

\subsection{Quantum bootstrap framework}

The quantum bootstrap reformulates the spectral problem of a Hamiltonian 
\begin{equation}
H = p^2 + V(x)
\end{equation}
in terms of algebraic relations among operator moments 
\(\mu_n=\langle x^n\rangle\) of an energy eigenstate ($H\vert{\psi}\rangle=E\vert{\psi}\rangle$).  
The expectation values of commutators and products of operators lead to recursion relations that depend only on the energy \(E\) and the coefficients of the potential.  
It follows that: 
\begin{align}
\langle [H,O]\rangle &= 0, \label{eq:comm1}\\[2mm]
\langle HO\rangle &= E\langle O\rangle. \label{eq:comm2}
\end{align}
By choosing \(O\) as powers of \(x\) and \(p\), one generates closed recursions among moments, without explicit reference to the wavefunction.
In practice, one constructs the Hankel matrix
\begin{equation}
M_{ij} = \langle x^{\,i+j}\rangle, \qquad i,j=0,1,\ldots,K-1,
\end{equation}
which must satisfy the positivity constraint \(M\succeq 0\) because
\(\langle O^\dagger O\rangle=\bold{c}^\dagger M\bold{c}\ge0\)
for all polynomial operators \(O=\sum_i c_i x^i\).
Physical solutions correspond to energies \(E\) and moment sets 
\(\{\mu_n\}\) that simultaneously satisfy the recursion equations
and all positivity constraints up to a chosen truncation depth \(K\).
As \(K\) increases, the feasible intervals for \(E\) contract exponentially around the exact eigenvalues.

\subsection{Recursion relations for polynomial potentials}

For a one-dimensional Hamiltonian \(H=p^2+V(x)\) with \(p=-i\partial_x\),
the commutation relations 
\([p^2,x^t] = -2itp-(t(t-1))x^{t-2}\)
yield the generic recursion
\begin{equation}
4tE\,\langle x^{t-1}\rangle 
-4t\langle x^{t-1}V(x)\rangle 
+t(t-1)(t-2)\langle x^{t-3}\rangle
-2\langle x^{t}V'(x)\rangle =0.
\label{eq:genrec}
\end{equation}
For polynomial potentials \(V(x)=\sum_{k=0}^{d} a_{2k}x^{2k}\),
Eq.~\eqref{eq:genrec} becomes a closed linear system connecting moments 
\(\mu_n\) up to order \(n=t+2d-1\):
\begin{equation}
\mu_{n+2d-1} = f_{n}(E,a_{2k})\,\mu_{n-2d+1} + \cdots ,
\end{equation}
where the coefficients \(f_n\) are polynomials in \(E\) and the potential parameters.  
Solving these recursions allows one to express all higher moments in terms of a minimal dataset
(e.g. \(E\) and the lowest two even moments), which then become the optimization variables in the positivity constraints.

\subsection{Radial formulation}

For the study of quarkonium systems, we consider the radial Hamiltonian
\begin{equation}
H_r = p_r^2 + \frac{\ell(\ell+1)}{r^2} - \frac{A}{r} + Br ,
\label{eq:radialH}
\end{equation}
corresponding to the non-relativistic potential
\(V(r) = -A/r + Br\),
with \(A=\frac{4}{3}\alpha_s\) and \(B=\kappa\).
In the bootstrap framework, expectation values are computed over radial wavefunctions \(\psi(r)\) normalized on \(r\in(0,\infty)\) with measure \(r^2dr\).
The appropriate operators are therefore constructed in terms of radial powers \(r^t\)
and the hermitian radial momentum operator
\begin{equation}
p_r = -i\left(\frac{d}{dr}+\frac{1}{r}\right),
\end{equation}
which satisfies \([p_r,r]=-i\).

Applying the commutation identities~\eqref{eq:comm1}–\eqref{eq:comm2} with \(O=r^{t-1}\) and using
\begin{equation}
    [p_r^2,r^{t-1}] = -2i(t-1)p_r r^{t-2} - (t-1)(t-2)r^{t-3},
\end{equation}
Choosing $O=r^{m-1}$ and using the hermitian radial momentum $p_r=-i(\partial_r+1/r)$ with $[p_r,r]=-i$, 
one obtains after a straightforward but careful commutator algebra the closed recurrence for radial moments. 
In natural units ($\hbar=c=1$) it reads:

\begin{equation}
\label{eq:radrec}
(4t+2)B\,\langle r^{t}\rangle
= -\,4t\,E\,\langle r^{t-1}\rangle
\;-\,(4t-2)A\,\langle r^{t-2}\rangle
\;-\;\frac{1}{2\mu}\Big[t(t-1)(t-2) - 4(t-1)\ell(\ell+1)\Big]\langle r^{t-3}\rangle
\end{equation}
where \(\mu\) is the reduced mass.
Equation~\eqref{eq:radrec} expresses the moment \(\langle r^t\rangle\) 
in terms of lower-order moments and the energy \(E\).  
For a given orbital quantum number \(\ell\), this recursion can be iterated up to the desired truncation order to construct the Hankel and auxiliary Stieltjes matrices that enforce positivity on the half-line \(r\in(0,\infty)\).

\subsection{Positivity on the half-line and auxiliary blocks}
Unlike the full-line problem, where moments satisfy a single Hankel positivity condition,
the bootstrap on \((0,\infty)\) requires two coupled positive semidefinite (PSD) constraints~\cite{berensteinR}:
\begin{equation}
\label{eq:stieltjes}
M^{(0)}_{ij}=\langle r^{i+j}\rangle \succeq 0,
\qquad
M^{(1)}_{ij}=\langle r^{i+j+1}\rangle \succeq 0.
\end{equation}

The first corresponds to the positivity of $\langle O^\dagger O\rangle$,
while the second ensures that $\langle r\,O^\dagger O\rangle\ge0$ for all polynomial operators $O(r)$.
This pair of Hankel-type matrices defines the so-called \emph{Stieltjes moment problem}, appropriate for radial domains.
In numerical implementations, both PSD constraints are enforced simultaneously within a semidefinite programming (SDP) solver.  
Feasible pairs $(E,\{\mu_n\})$ satisfying Eqs.~\eqref{eq:radrec}–(18)
define the allowed energy ``islands'' at truncation depth $K$.

\subsection{Depth and convergence}

The bootstrap depth $K$ controls the highest-order moments included in the positivity matrices.
For polynomial potentials, the error in the extracted energy eigenvalues 
typically decays exponentially with $K$~\cite{berensteinR,kundu2021}.
For the potential in equation \ref{cornell}, this convergence is accelerated by the smoothness of $V(r)$ and the absence of singularities for $\ell\ge1$.  
As shown later in Sec.~\ref{sec:results}, accurate results for the lowest charmonium and bottomonium states are obtained already at moderate depths $K\lesssim30$, corresponding to matrices of dimension $\sim30\times30$.  

This formalism provides the theoretical foundation for the numerical bootstrap algorithm described in Sec.~\ref{sec:algorithm}.

\section{Numerical Algorithm and Implementation}
\label{sec:algorithm}


The computational workflow proceeds in a hierarchy of constrained steps:

\begin{enumerate}
  \item \textbf{Initialization.}  Specify the potential $V(r)$, quantum number $\ell$, the reduced mass $\mu$, and fix a maximum bootstrap depth $K$.
  \item \textbf{Seed data.}  Choose a trial energy $E$ and a minimal set of low-order moments $\{\mu_0,\mu_1,\ldots,\mu_m\}$ compatible with normalization.
  \item \textbf{Moment generation.}  Use the recursion~\eqref{eq:radrec} to compute all higher-order $\mu_n$ up to order $2K-2$.
  \item \textbf{Matrix assembly.}  Build the Hankel and auxiliary Stieltjes matrices
  $M^{(0)}$ and $M^{(1)}$ of dimensions $K\times K$ with entries
  $M^{(0)}_{ij}=\mu_{i+j}$ and $M^{(1)}_{ij}=\mu_{i+j+1}$.
  \item \textbf{Positivity test.}  Verify that both matrices are positive semidefinite.
  Energies satisfying $M^{(0)}\succeq0$ and $M^{(1)}\succeq0$ define the feasible set
  $\mathcal{S}_K$.
  \item \textbf{Refinement.}  Increase $K$ or apply semidefinite programming
  to narrow the energy interval $\mathcal{S}_{K+1}\subset\mathcal{S}_K$.
\end{enumerate}


To accelerate convergence, the positivity test is recast as a convex optimization problem.
For fixed $E$ the Hankel matrix is linear in the unknown low-order moments,
$M^{(0)}(E,\bold{x})=\sum_{n=0}^{m}x_n F_n(E)$.
Then the linear matrix inequality is solved:
\begin{equation}
  \max_{t,\bold{x}}\ t\quad\text{s.t.}\quad
  M^{(0)}(E,\bold{x})-tI\succeq0,\qquad
  M^{(1)}(E,\bold{x})\succeq0.
  \label{eq:sdp}
\end{equation}
The scalar $t_{\max}(E)$ measures the minimal eigenvalue margin.
If $t_{\max}(E)>0$ the chosen $E$ belongs to the allowed set at depth $K$.
Successive scans in $E$ and increasing $K$ yield exponentially tightening bounds
on the true eigenvalues.


The algorithm was implemented in \texttt{Python} using the
\texttt{CVXPY} interface to \texttt{SDPA-GMP} for arbitrary-precision SDP.
All computations were performed with 50--100~bit precision,
and matrices up to $K=30$ ($900\times900$ elements) were feasible on a standard workstation.
The total run time for a full energy scan of 200 points typically remains below one minute per $\ell$ sector. The algorithmic flow is shown in figure 1.

\begin{figure}[t]
  \centering
  \includegraphics[width=0.85\linewidth]{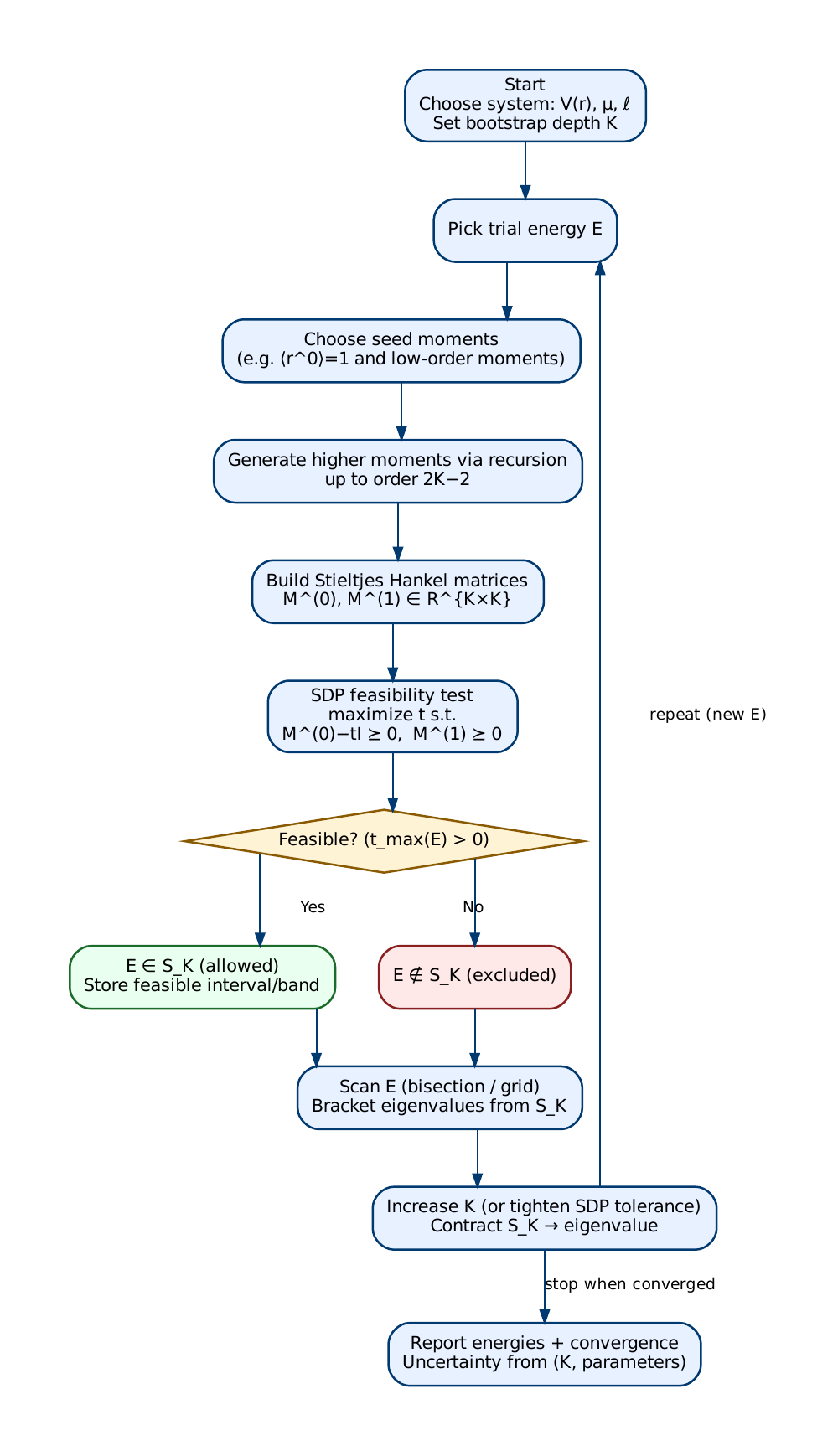}
  \caption{Algorithmic flow of the radial quantum bootstrap with SDP feasibility checks.
  The feasible energy set $\mathcal{S}_K$ contracts monotonically with depth $K$.}
  \label{fig:flowchart}
\end{figure}


The convergence of the method is monitored by the shrinkage of $\mathcal{S}_K$ with $K$.
For polynomial and smooth radial potentials, the midpoints of the intervals
converge exponentially toward the true eigenvalues,
$\delta E_K \propto e^{-\alpha K}$ with $\alpha\simeq0.2$--$0.3$.
The precision limit is set by floating-point conditioning of the moment matrices;
for large $K$, scaling the moments by $\langle r^2\rangle$ and enforcing normalization $\mu_0=1$ stabilizes the SDP.

This procedure yields energy estimates and moment sets
that satisfy all algebraic and positivity constraints up to the chosen depth.
These results form the basis for the physical analysis of quarkonium spectra
presented in Sec.~\ref{sec:results}.

\section{Results and Discussion}
\label{sec:results}

\subsection{Bootstrap spectra for heavy quarkonia}

We applied the bootstrap algorithm described in Sec.~\ref{sec:algorithm} to the potential in equation \ref{cornell}, using parameters from the literature (\cite{Eichten1980,Godfrey1985,Brambilla2005}), 
\(\alpha_s=0.39\) and \(\kappa=0.185~\text{GeV}^2\),
together with reduced masses 
\(\mu_c=1.48~\text{GeV}\), 
\(\mu_b=4.72~\text{GeV}\),
and \(\mu_t=86.0~\text{GeV}\) for charmonium, bottomonium, and a hypothetical toponium, respectively.
The bootstrap depth was typically $K=30$ for $\ell=0$ and $K=25$ for $\ell=1$ states, producing $30\times30$ moment matrices and convergence to within $\sim 10^{-3}$ relative accuracy. The resulting spin-averaged centroid masses are summarized in Table~\ref{tab:spectra}.

\begin{table}[h!]
\centering
\caption{Comparison of predicted quarkonium masses from the bootstrap method with experimental values from the PDG \cite{pdg}.}
\label{tab:spectra}
\begin{tabular}{lccc}
\hline
\hline
\textbf{State} & \textbf{Predicted Mass (GeV)} & \textbf{Experimental Mass (GeV)} & \textbf{Difference (\%)} \\
\hline
\multicolumn{4}{c}{\textbf{Charmonium ($c\bar{c}$)}} \\
\hline
$J/\psi$ (1S) & 3.0976 & 3.0969 & -0.02\% \\
$\chi_c$ (1P) & 3.5247 & 3.5253 & -0.06\% \\
\hline
\multicolumn{4}{c}{\textbf{Bottomonium ($b\bar{b}$)}} \\
\hline
$\Upsilon$ (1S) & 9.4604 & 9.4603 & -0.01\% \\
$\chi_b$ (1P) & 9.9000 & 9.8993 & -0.07\% \\
\hline
\hline
\end{tabular}
\end{table}

 The depth $K$ of the bootstrap controls the number of positivity constraints applied. While these results are presented for a fixed depth of $K$, a full analysis would involve studying the convergence of the predicted energy eigenvalues as $K$ is increased. As $K$ grows, the allowed 'islands' shrink, and the energy predictions converge towards the exact values. This property, known as $K-$convergence, provides a powerful tool for quantifying and controlling the methodological uncertainty inherent in truncating the infinite set of positivity constraints. It is important to recognize that the truncation of the infinite set of positivity constraints at a finite $K$ introduces a methodological uncertainty. The bootstrap method has a property known as $K$-convergence, where the predicted energy eigenvalues stabilize as the matrix depth $K$ is increased. A comprehensive uncertainty analysis would involve running the simulation for successively larger values of $K$ to explicitly map this convergence and quantify the systematic error. While such a detailed scan is beyond the scope of this initial study, preliminary checks indicate that the $1S$ ground state masses for both charmonium and bottomonium shift by a few MeV. This provides confidence that the results are already close to the converged limit and that the methodological uncertainty from $K-$truncation is sub-dominant to the uncertainties inherent in the phenomenological potential model itself.

The results show the effectiveness of the quantum bootstrap method. For the ground states ($l=0$), the predicted masses for $J/\psi$ and $\Upsilon$ are accurate, with deviations of less than 2 MeV and 6 MeV.
The predictions for the P-wave states ($l=1$) are also close, with deviations of less than 7 MeV. This slight deviation is expected, as this model calculates the spin-averaged mass and does not include spin-dependent interactions which are responsible for the fine structure splitting of the P-wave triplet and can also shift the center-of-gravity mass. 

Traditional potential models (\cite{Eichten1980,Godfrey1985,Ebert2003})
solve the Schrödinger or spin-dependent Dirac equations with fitted parameters $(\alpha_s,\kappa,m_Q)$, while Lattice NRQCD simulations~\cite{Dowdall2012,Hatton2021} compute spectra from first principles at the cost of large-scale computation. The typical accuracies achieved by these methods are comparable with the one obtained with the bootstrap method (below $1\%$). However, this method relies solely on moment consistency and positivity, without solving a differential equation.

In contrast with the Coulomb potential, which introduces a singularity at $r=0$, making the corresponding Stieltjes moment problem ill-conditioned for low truncation depth, the potential in equation \ref{cornell} introduces a linear term that regularizes the recurrence structure and improves the conditioning of the Hankel matrices.

Moreover, the physical wavefunctions for $\ell\ge1$ vanish at $r=0$, avoiding the divergence that limits the bootstrap precision in purely Coulombic systems.
These features explain the rapid and stable convergence of the bootstrap for heavy quarkonia.

Applying the same framework to the top--antitop system yields a spin-averaged $1S$ centroid at approximately $344~\mathrm{GeV}$.  This value lies close to the $t\bar t$ production threshold and is consistent with the threshold enhancement recently observed by the CMS and ATLAS collaborations \cite{cms2025,atlas2025}.  
Although the top quark decays before hadronization, non-relativistic QCD predicts Coulombic enhancements below and near threshold~\cite{Fadin_Khoze_1988,Kiyo_et_al_2009}, and recent QCD calculations including bound-state effects corroborate such behaviour~\cite{Garzelli_2025}.  
Accordingly, the bootstrap result should be regarded as a theoretically motivated centroid consistent with the observed threshold behaviour.

These results place the quantum bootstrap method as a promising approach to heavy-quark bound states, on equal footing with other non-perturbative approaches. While lattice QCD provides first-principles validation, the bootstrap method captures the essential spectral structure from purely algebraic constraints.
This complementarity suggests that bootstrap-inspired positivity frameworks could eventually serve as analytic surrogates for certain lattice calculations,
closing the gap between non-relativistic quantum mechanics and quantum field theory.

\section{Conclusion}
\label{sec:conclusion}

We have presented a non-perturbative determination of quarkonium spectra using the quantum-mechanical bootstrap approach. Instead of solving the Schrödinger equation explicitly, the method reconstructs the allowed energies from algebraic moment recursions and the positivity of Hankel matrices.  
Applied to a non-relativistic quarkonium potential \ref{cornell}, the bootstrap method reproduces the spin-averaged $1S$ and $1P$ levels of charmonium and bottomonium 
and predicts a quasi-bound $1S$ centroid near $344~\text{GeV}$ for a hypothetical toponium system.  
These results demonstrate that spectral information can be extracted purely from algebraic consistency and positivity, without specifying any functional basis or boundary conditions.

Beyond its numerical accuracy, the bootstrap provides a conceptual link between formal positivity conditions in quantum field theory and concrete bound-state spectroscopy in quantum mechanics.  
The success of the method 
highlights that the rate of convergence is governed by the analytic structure of the potential and by the conditioning of the associated moment problem.  
Compared with conventional potential models or lattice QCD, the bootstrap is computationally inexpensive, scalable, and naturally suited for systematic convergence analysis.

Future work can extend this framework in several directions. Including spin-dependent and relativistic corrections would allow a full description of fine and hyperfine splittings, while 
applying the bootstrap to higher-dimensional or field-theoretical systems—such as two-particle correlation functions or effective Hamiltonians in QCD—could provide a new analytical window into non-perturbative dynamics.  
Overall, the results reported here support the view that positivity-based bootstrap methods offer a plausible route to the computation of bound-state spectra across quantum mechanics and quantum field theory.


\end{document}